\documentstyle[11pt,newpasp,twoside,epsf]{article}
\def\edcomment#1{\iffalse\marginpar{\raggedright\sl#1\/}\else\relax\fi}
\reversemarginpar

\begin{document}
\title{Far-red spectroscopy of peculiar stars and the GAIA mission}
 \author{Ulisse Munari}
\affil{Osservatorio Astronomico di Padova, Sede di Asiago,\\ 36012 Asiago (VI), 
Italy,  ~~~~~{\rm (}{\tt munari@pd.astro.it}{\rm )}}

\begin{abstract}
The coming GAIA Cornerstone mission by ESA will provide micro-arcsec
astrometry, $\sim$10 bands photometry and far-red spectroscopy for a huge
number of stars in the Galaxy (10$^9$). GAIA spectroscopy will cover the
range 8480--8740 \AA\, which includes the CaII triplet and the head of the
Paschen series. In this paper we address the diagnostic potential of this
wavelength range toward detection of peculiar stars.
\end{abstract}

\section{Introduction}

The large impact that the Hipparcos astrometric mission by ESA had on many
fields of astrophysics is well known to the community. This is even more
remarkable by considering that the Hipparcos observations were complete to
just $V\sim$8 mag and the horizon for astrometric errors less than 10\%
($R_{10\%}$) was limited to 0.1 kpc, i.e. the solar neighborhood.
Nevertheless, Hipparcos data have been the main driver or at least a
contributor to more than 1900 papers since the release to the community of
the mission data in 1997.

Hipparcos was still flying when the European community begun openly speaking
of its successor, with ideas already well in focus about GAIA by the time of
the Cambridge 1995 ESA colloquium on the future of astrometry in space (ESA
SP-379). Since then, GAIA has been at the center of a continent-wide effort
dealing with its science goals and the technical design, culminated with the
formal mission approval in the fall of 2000. Perryman et al. (2001, and
references therein) provides an useful introduction to GAIA.

Before GAIA (http://astro.estec.esa.nl/GAIA/) which launch is scheduled for
not later than 2012, two other survey astrometric missions could fly if
their current financial problems will be eventually overcome (cf. Table~1):
the German DIVA (http://www.ari.uni-heidelberg.de/diva/) and the USA mission
FAME (http://www.usno.navy.mil/FAME/), both with astrometric goals
intermediate between those of Hipparcos and GAIA. Finally, the technological
demonstrator for interferometry in space SIM (http://sim.jpl.nasa.gov/),
base-lined by NASA for a launch in 2009, will obtain micro-arcsec astrometry
of a preselected limited sample of objects. While all satellites will perform
photometry in parallel with astrometry, only GAIA will also collect spectra
with the main aim of deriving radial velocities and therefore to measure the
6$^{th}$ component of the phase-space (the other five being provided by
astrometry).

\begin{table}[!t]
\caption{Comparison between the performances of Hipparcos, DIVA, FAME, 
SIM and GAIA astrometric missions (courtesy F.Mignard).}
\begin{tabular}{lccccc}
&\\
\tableline
&Hipparcos&DIVA&FAME&SIM&GAIA\\
\tableline
Mission Type & scanning & scanning & scanning & pointing & scanning \\
Input Catalogue & Yes & No & Yes & Yes & No \\
V$_{completness}$ (mag) & 8 &12.5  &14  & -- & 20 \\
N. of objects & 118,218 & $3\cdot 10^7$ & $4\cdot 10^7$ 
              & 20,000 & $1\cdot 10^9$ \\
$\sigma_\pi$ ($\mu$as)& 1000$_{(V=9)}$ & 200$_{(V=9)}$ & 50$_{(V=9)}$ 
              & 4$_{(V=16)}$ & 10$_{(V=15)}$\\
R$_{10\%}$ (kpc) & 0.1 & 0.5 & 2 & 25 & 10 \\
Spectra and RV & No & No & No & No & Yes \\    
\tableline\tableline
\end{tabular}
\end{table}

\section{GAIA spectroscopy}

GAIA will survey the whole sky by spinning every three hours around an axis
that in turn will precess around the Sun. For a given star during a single
spin rotation, GAIA will collect one astrometric measurement, one spectrum
and one photometric reading in each of $\sim$10 bands. During the planned
5-year mission lifetime, while scanning the sky GAIA will visit each star
about 100 times. Therefore about 100 astrometric positions, 100 spectra and
100 readings in each photometric band will be recorded for each target star
during the mission life-time.

The stars will cross GAIA field of view for a time interval fixed by the
satellite spin rate, and this sets the equivalent {\it exposure time} for
the CCDs on the focal plane (which are read in TDI mode), i.e. an exposure
time equal for all stars independently from their magnitude. Assuming a
resolving power of $\sim$10,000 and a dispersion of 0.5~\AA/pix for GAIA
spectra (still to be finalized), and the current optical design and
efficiencies, Table~2 lists the magnitude of stars producing spectra of
S/N=100, 30 and 10.

\begin{figure}
\plotone{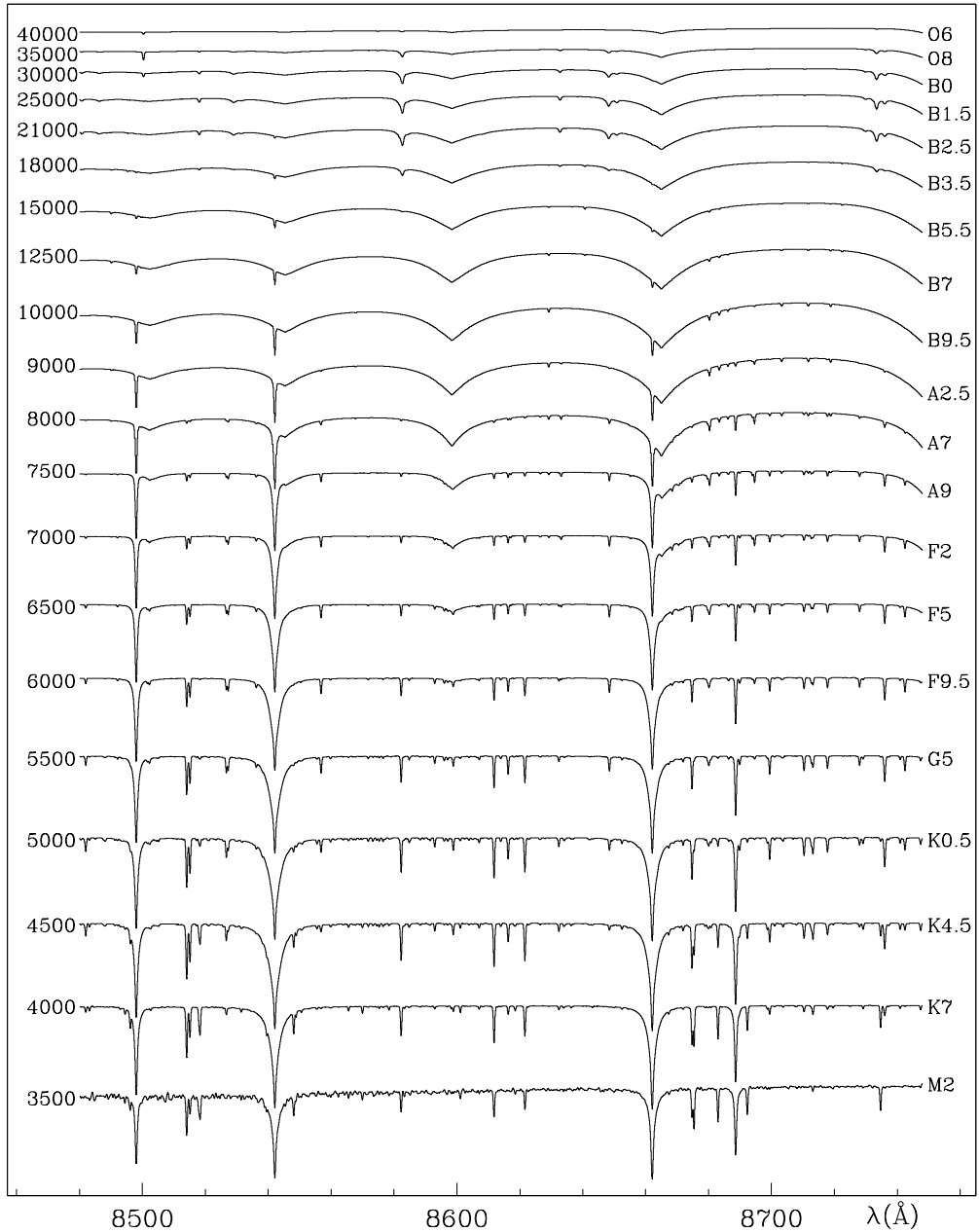}
\caption{Sequence of synthetic spectra (from Munari \& Castelli 2000,
Castelli \& Munari 2001) illustrating the variations along the main sequence
($T_{eff}$ in K on the left and corresponding spectral type for main
sequence stars on the right) for moderately metal poor stars
([Z/Z$_\odot$]=$-$0.5). All spectra are on the same ordinate scale, only
displaced in their zero-points. A detailed mapping of the MKK classification
system with real spectra is provided in the atlas by Munari and 
Tomasella (1999).}
\end{figure}

\begin{figure}
\plotone{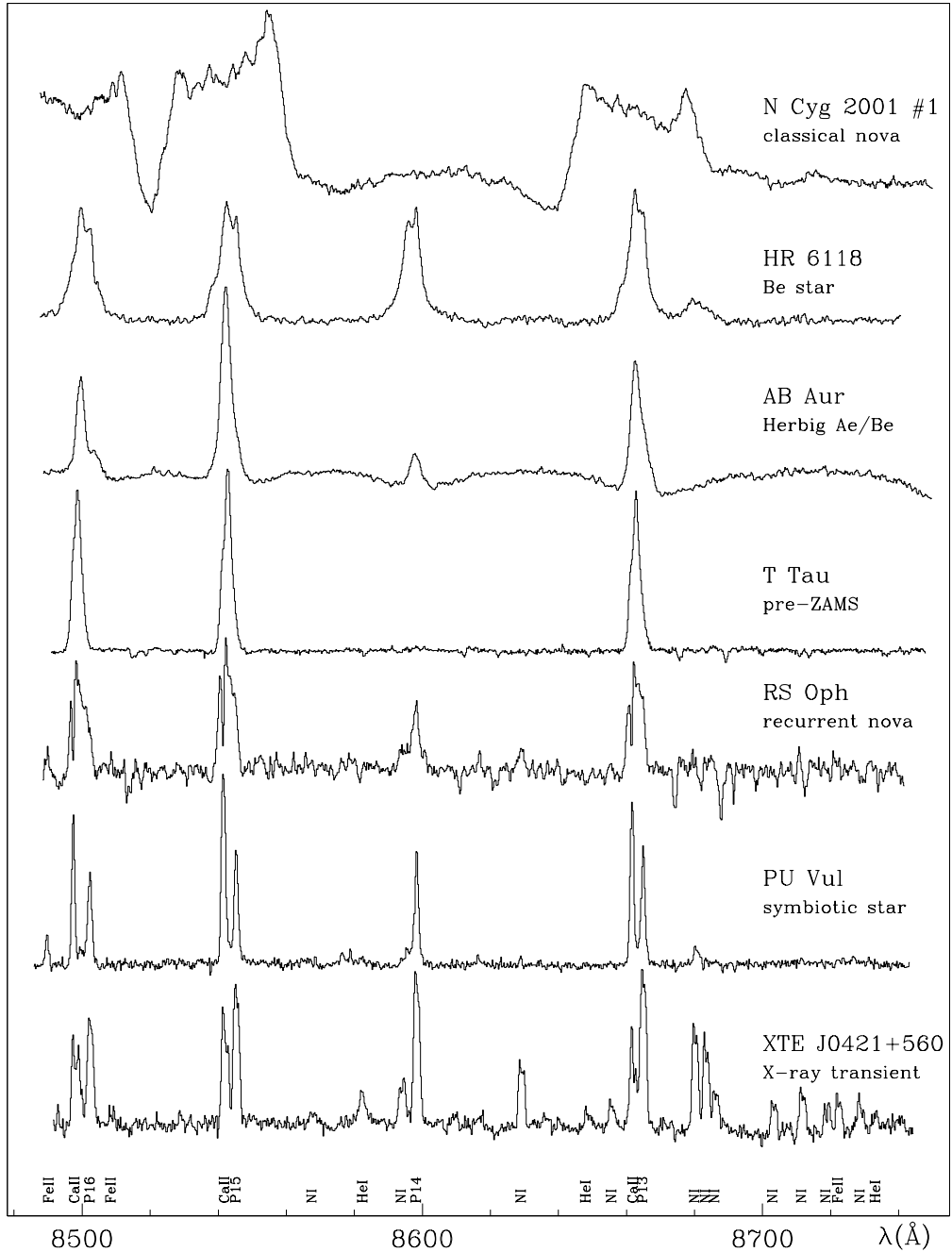}
\caption{Examples of far-red spectra of peculiar objects dominated by
emission lines (see identification at bottom). Spectra obtained with the
Asiago Echelle+CCD spectrograph. The expected GAIA spectra should resemble
those shown here.}
\end{figure}

\begin{figure}
\plotone{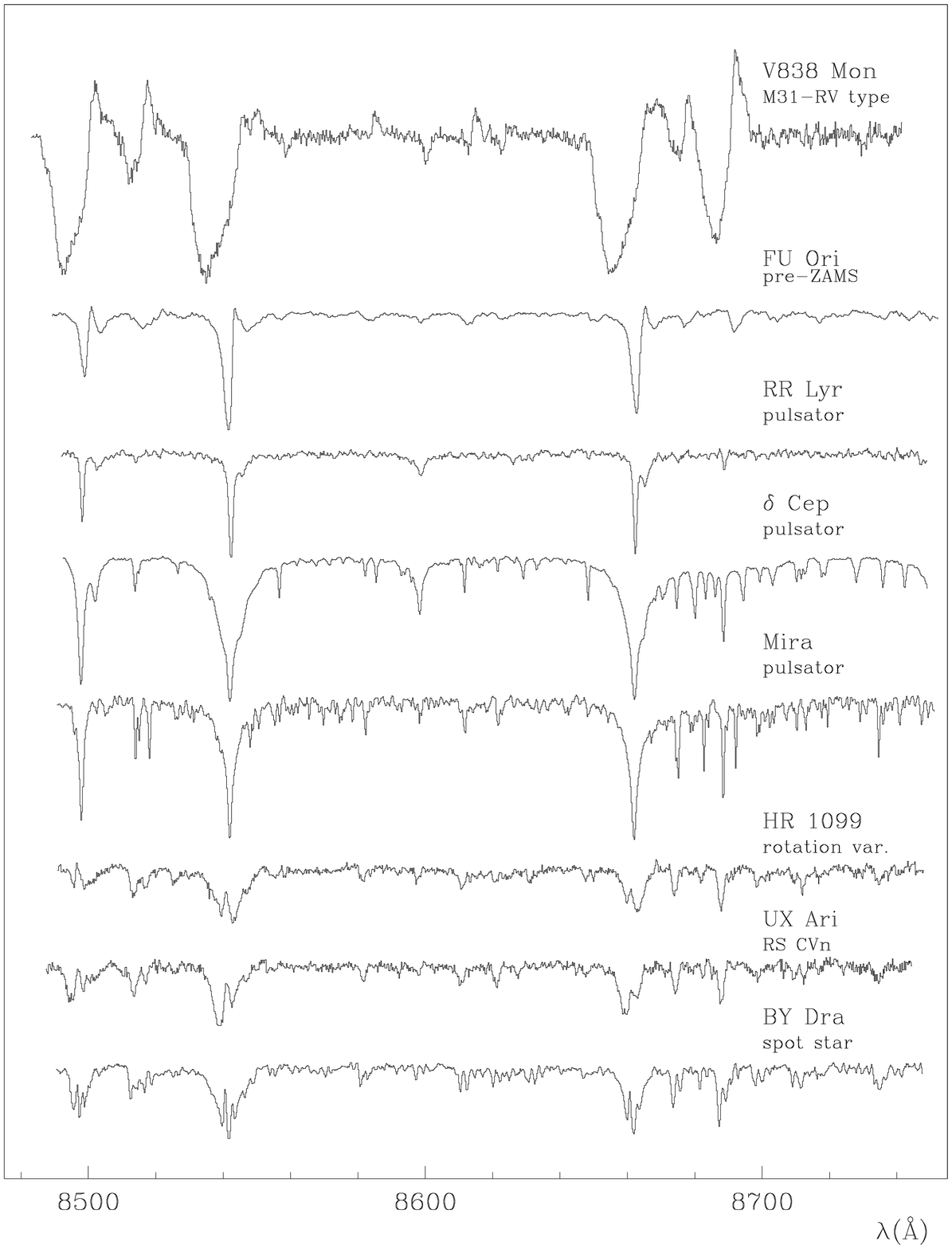}
\caption{Additional examples of far-red spectra of peculiar objects, this
time dominated by absorptions (mainly from CaII, FeI, TiI, MgI, SiI, MnI,
SI, CN and TiO). Note the P-Cyg CaII profiles in V838~Mon and FU~Ori, and
the emission components within the CaII absorption profiles for the stars
with active photospheres (HR 1099, UX Ari and BY Dra). Spectra of prototype
pulsating stars are also shown.}
\end{figure}

GAIA will record spectra over the 8480-8740 \AA\ range, covering the CaII
triplet, the head of the Paschen series, NI multiplets \#1 and \#8, many
lines of FeI, TiI, MgI, SiI, MnI, SI and the CN and TiO molecules. Figure~1
show the progression of spectral changes in the GAIA wavelength region along
the MKK classification system for normal stars. The classification potential
is excellent, particularly for cool stars (G-K types being the dominant
population among field stars at the faint magnitudes reached by GAIA).
General introduction to GAIA spectroscopy is provided by Munari (1999) and
Munari (2002), an evaluation of GAIA performances on eclipsing binaries is
given by Munari et al. (2001a) and Zwitter (this volume), and the
accuracy of GAIA radial velocities has been investigated by Munari et al.
(2001b), Katz et al. (2002) and Zwitter (2002).

Figures~2 and 3 present a sample of spectra of peculiar and exotic objects,
extracted from a much larger survey performed with the Echelle+CCD
spectrograph at the 1.82 m telescope in Asiago, operated in GAIA modality
(Munari et al. 2002, A\&A, to be submitted).

The various types of peculiar stars well distinguish among themselves and
with respect to normal stars by the presence of emission lines of different
species and intensity, and characterized by profiles of different shape and
width. In cool peculiar stars the emission in the CaII dominates (cf.
T~Tau), and as the temperature rises Paschen lines grow in strength while
CaII decreases (cf. AB~Aur). As the temperature further increases (cf. HR
6118) CaII vanishes and Paschen lines reach peak intensities. Objects with
highly stratified and distinct emission regions (cf. PU~Vul and
XTE~J0421+560) present a wide range of emission lines characterized by quite
different excitation conditions, including CaII, Paschen, HeI, NI and FeII.
Fast expansion and ejecta clumpiness in novae (cf. N Cyg 2001 \#1) is well
traced in the CaII profiles close to maximum and by Paschen lines at later
phases, while conspicuous mass loss (cf. V838~Mon) manifests in wide P-Cyg
profiles. The presence of active regions on the surface of cool stars is
revealed by emission components within the CaII absorption profiles (cf.
HR~1099, UX~Ari and BY~Dra), which change in position and intensity traces
the rotation of the stars and the life-time of the hot spots.

\begin{table}[!t]
\caption{The table provides the magnitudes of the stars that produce GAIA
spectra with S/N=100, 30 or 10 (per pixel, at 0.5~\AA/pix dispersion and
10,000 resolving power) per single passage (values to the left) and
mission-averaged (values to the right). The magnitudes are computed for the
Cousins' $I$ band, which covers the wavelength range of GAIA spectra. The
corresponding $V$ magnitudes are listed for F5 ($V-I_{\rm C}$=+0.51), G5
($V-I_{\rm C}$=+0.73) and K5 ($V-I_{\rm C}$=+1.34) unreddened main sequence
stars (G-K stars are the dominant spectral types of field stars at the faint
magnitudes reached by GAIA).}
\begin{tabular}{r|rrrr|rrrr}
\multicolumn{9}{c}{}\\
\tableline
S/N & \multicolumn{1}{c}{$I$} & $V_{\rm F5 V}$ &  $V_{\rm G5 V}$ & $V_{\rm K5 V}$
&\multicolumn{1}{c}{$I$} & $V_{\rm F5 V}$ &  $V_{\rm G5 V}$ & $V_{\rm K5 V}$ \\
\tableline
100  &   8.66  &    9.17 &  9.39 & 10.00 &   13.56  &   14.07 & 14.29 & 14.90 \\
 30  &  11.27  &   11.78 & 12.00 & 12.61 &   16.07  &   16.58 & 16.80 & 17.41 \\
 10  &  13.51  &   14.02 & 14.24 & 14.85 &   18.21  &   18.72 & 18.94 & 19.55 \\
\tableline\tableline
\end{tabular}
\end{table}


\begin{references}
\reference Castelli, F., Munari, U. 2001, A\&A 366, 1003
\reference Katz, D., Viala, Y., Gomez, A., Morin, D. 2002, in `GAIA: A European 
    Space Project', O. Bienaym\'{e} and C. Turon ed.s, (Paris: EDP Sciences), 64
\reference Munari, U. 1999, in ESA GAIA Workshop, V. Strayzis ed., 
    Balt. Astron. 8, 73
\reference Munari, U. 2002, in `GAIA: A European Space Project', O. Bienaym\'{e} 
    and C. Turon ed.s, (Paris: EDP Sciences), 39
\reference Munari, U., Tomasella L.  1999, A\&AS 137, 521
\reference Munari, U., Castelli F.  2000, A\&AS 141, 141
\reference Munari, U., Tomov, T., Zwitter, T. et al. 2001a, A\&A, 378, 477
\reference Munari, U., Agnolin, P., Tomasella, L. 2001b, Balt. Astron. 10, 613
\reference Perryman, M.A.C., de Boer, K.S., Gilmore, G. et al. 2001, A\&A 369, 339
\reference Zwitter, T. 2002, A\&A 386, 748
\end{references}
\end{document}